\documentclass[10pt, conference, letterpaper]{IEEEtran}
\IEEEoverridecommandlockouts

\usepackage{amsmath,amsthm,amssymb,amsfonts}
\usepackage{cite}
\usepackage{url}
\usepackage{algorithmic}
\usepackage{graphicx}
\usepackage{subcaption} 
\usepackage{textcomp}
\usepackage{xcolor}
\usepackage{bm}
\usepackage{multirow}

\newtheorem{assmp}{Assumption}

\newtheorem{lemma}{Lemma}

\newtheorem{prop}{Proposition}

\newtheorem{example}{Example}

\newtheorem{defn}{Definition}
\usepackage{lipsum}

\def\BibTeX{{\rm B\kern-.05em{\sc i\kern-.025em b}\kern-.08em
    T\kern-.1667em\lower.7ex\hbox{E}\kern-.125emX}}
\begin{document}

\title{Lightweight Federated Learning with Differential Privacy and Straggler Resilience}


%
\author{Shu Hong, Xiaojun Lin and Lingjie Duan
\IEEEcompsocitemizethanks{
S. Hong and L. Duan are with Singapore University of Technology and Design, Singapore (email: shu\_hong@mymail.sutd.edu.sg, lingjie\_duan@sutd.edu.sg).
X. Lin is with Purdue University, West Lafayette, IN, USA (email: linx@ecn.purdue.edu).
}
}




\maketitle
\begin{abstract}
Federated learning (FL) enables collaborative model training through model parameter exchanges instead of raw data. To avoid potential inference attacks from exchanged parameters, differential privacy (DP) offers rigorous guarantee against various attacks. However, conventional methods of ensuring DP by adding local noise alone often result in low training accuracy. 
Combining secure multi-party computation (SMPC) with DP, while improving the accuracy, incurs high communication and computation overheads as well as straggler vulnerability, in either client-to-server or client-to-client links.
In this paper, we propose LightDP-FL, a novel lightweight scheme that ensures provable DP against untrusted peers and server, while maintaining straggler resilience, low overheads and high training accuracy. 
Our scheme incorporates both individual and pairwise noise into each client's parameter, which can be implemented with minimal overheads. 
Given the uncertain straggler and colluder sets, we utilize the upper bound on the numbers of stragglers and colluders to prove sufficient noise variance conditions to ensure DP in the worst case. Moreover, we optimize the expected convergence bound to ensure accuracy performance by flexibly controlling the noise variances.
Using the CIFAR-10 dataset, our experimental results demonstrate that LightDP-FL achieves faster convergence and stronger straggler resilience compared to baseline methods of the same DP level.
\end{abstract}
%



\section{Introduction}
\label{Sec:Introduction}


\subsection{Background and Motivations}

Federated learning (FL) is an emerging distributed machine learning paradigm that enables collaborative training of models using private data, without the need to transmit local data to a central server. In a typical FL scheme, each client computes the gradient of the local loss function over his own dataset, and transmits only the gradient or corresponding parameter to the server. 
However, keeping data locally does not inherently secure privacy, as adversaries can potentially reconstruct raw private data or perform membership inference attacks from the exchanged parameters alone, as demonstrated by recent studies  \cite{zhao2020idlg, wang2022protect}.
%

To address the privacy concerns related to gradient exposure, recent research has focused on enhancing FL mechanisms with differential privacy (DP) \cite{dwork2014algorithmic}. DP provides a rigorous guarantee that inputs remain indistinguishable from all possible observations, including both individual and aggregated global outputs. 
Implementing DP in FL \cite{abadi2016deep, wei_federated_2020,yang2023privatefl} typically involves introducing random noise to parameters to mask sensitive information. Although this scheme offers a strong mathematical defense against various attacks\footnote{Notably, while some recent works  \cite{wang2022protect,wang2023more} propose defensive mechanisms against specific gradient-based attacks, 
they may not address all potential vulnerabilities. In contrast, the DP guarantees discussed in this paper are designed to be effective against any attack scenario.}, it involves the addition of local noise for each client. Such local noise accumulates during the server-side aggregation, leading to a significant increase in overall noise and, consequently, substantial losses in training accuracy.
      
Since the server only needs the sum of the parameters, there is an opportunity to combine DP with other techniques such as secure multi-party computation (SMPC), so that only a small amount of additional noise is needed to ensure DP.
In SMPC schemes, individual updates from distributed clients to the server are fully secure based on coded computation  \cite{jahani2023swiftagg+}, pairwise encryption \cite{mugunthan2019smpai,bonawitz2017practical}, or homomorphic encryption (HE)  \cite{paillier1999efficient,damgaard2001generalisation}.
Note that these SMPC mechanisms alone do not ensure DP, because the revealed sum at the untrustworthy server side is unperturbed, which will disclose sensitive information from the dataset. Nonetheless, it is not hard to add an additional, but small, amount of random noise to each client’s parameter, so that the total amount of noise added to the sum is sufficient to preserve DP \cite{jiang2024dordis,truex2019hybrid}. The amount of noise added in this way is usually much lower (by a factor $\frac{1}{N}$, where $N$ is the number of clients), compared to the vanilla case of only adding the random noise without using SMPC.


Despite the promising benefits of combining SMPC with DP, each of these three types of SMPC schemes have shortcomings in their vulnerability to stragglers and/or high computation/communication overheads.


\begin{table*}[!ht]
	\centering
	\caption{Comparison of Existing Privacy Mechanisms and the Proposed Scheme for Federated Learning}
	\begin{tabular}{|p{4.7cm}|p{1.2cm}|p{2.1cm}|p{3.4cm}|p{1.1cm}|}
		\hline
		Privacy mechanism for FL & DP-private 
		& Client-to-server /client-to-client straggler-resilient
		& Communication/computation overhead & Accuracy  \\ \hline
	Vanilla local DP noise-adding (e.g.,   \cite{abadi2016deep}) & yes 
		& yes 
		& low & low  \\ \hline
	Homomorphic encryption+DP noise \cite{truex2019hybrid}  & yes  
		& yes 
		& high & high  \\ \hline
	Coded computation  (e.g., 	SwiftAgg+  \cite{jahani2023swiftagg+})  & no\footnotemark[2] 
		& no 
		& medium & high  \\ \hline
Pairwise encryption (e.g., \cite{mugunthan2019smpai})  & no\footnotemark[2] 
		& no 
		& low & high  \\ \hline
		SecAgg   \cite{bonawitz2017practical} & no\footnotemark[2]
		& no 
		& high & high    \\ \hline 
		Dordis (SecAgg+DP noise)  \cite{jiang2024dordis}  & yes 
		& no 
		& high & high  \\ \hline
		Our scheme (LightDP-FL) & yes 
		& yes 
		& low & high \\
		\hline 
	\end{tabular}
	\label{Table:literature}
\end{table*}

\begin{itemize}
\item 
Straggler issue: Straggling clients can potentially result in substantial delays or even failure in FL systems. For example, the pairwise encryption scheme  \cite{mugunthan2019smpai}
is particularly sensitive to
stragglers due to large key terms. 
Recent works \cite{jahani2023swiftagg+,jiang2024dordis,bonawitz2017practical} that claim to be straggler-resilient introduce additional transmissions and rely heavily on client-to-client secret sharing in each round for full information recovery. 
Their hidden assumption that all clients can consistently communicate with their peers is impractical. Consequently, these schemes exhibit vulnerability to stragglers in many client-to-client and client-to-server links in the course of the transmission and decryption phases (see detailed discussions in Section \ref{Sec:Related Work}). This may even result in system breakdown or divergence.

\item 
Computation/communication overheads: These overheads are significant bottlenecks in FL, particularly at the distributed client side. For example, the employment of HE scheme  \cite{truex2019hybrid} introduces high communication loads due to the large size of ciphertexts and an additional server-to-client round for partial decryption. High computation overhead arises from the robust cryptographic techniques and validation operations required. 
Another notable SMPC scheme, SecAgg  \cite{bonawitz2017practical}, also involves the client-to-server transmission of large-size encrypted parameters.
In the presence of stragglers, this scheme incurs extra overheads for client-to-client secret sharing {\textit{in every round}} due to additional reconstruction rounds. Without frequent secret sharing, if a client drops out once and his peers disclose the pairwise masks for full information recovery, then an adversary can easily infer his ground-truth parameters.

\end{itemize}

%
%

Table \ref{Table:literature} summarizes the challenges faced by existing literature.
In summary, it remains an open challenge to {\textit{develop an efficient and lightweight FL mechanism that offers strict DP guarantees against untrustworthy colluding peers and server, while maintaining resilience against stragglers and incurring low computation/communication overhead}s}.

\footnotetext[2]{Note that it is not hard to further add a small amount of random noise for these SMPC schemes to ensure DP.}
\setcounter{footnote}{2}

\subsection{Contributions}
In this paper, we present a novel FL mechanism, LightDP-FL, that is not only differentially private, but also lightweight and straggler-resilient. Note that the existing SMPC works aim for perfect information-theoretic privacy for clients (i.e., from aggregated parameters, adversaries cannot infer individual client information). 
This level of privacy often comes with significant communication and computational burdens and is vulnerable to unstable client-to-client and client-to-server links.
However, the server-side aggregation only requires DP-level protection. 
Then we naturally ask whether we really need the individual client-to-server updates to be fully safe with those complicated secure aggregation protocols. 
Motivated by this idea, we propose a lightweight scheme to deal with the open question above and provide a rigorous DP guarantee against colluders, stragglers, and unknown attacks in challenging environments.

%

Since our goal is to ensure DP, our key idea is to replace the complicated SMPC with simple pairwise masking schemes. Specifically, we leverage an individual noise term $\bm{n}_i$ and pairwise noise terms $\bm{r}_{ij}$, all following Gaussian distributions, to mask local outcomes for each client $i$. 
During the aggregation, the pairwise masks $\bm{r}_{ij}$'s are canceled out with no stragglers. 
When there are stragglers, we properly control the variances of $\bm{r}_{ij}$'s and $\bm{n}_i$'s to minimize the degradation of accuracy, while retaining sufficient DP guarantees. 
As a result, even if some clients straggle, as long as their corresponding pairwise noise terms have a sufficiently low variance, the accuracy can approach the optimal level. This is the key reason that LightDP-FL can retain the low computation/communication overheads, while being straggler-resilient at the same time.
Despite its potential benefits, designing provable DP-FL is challenging, as the pairwise noise terms are released by more than one client.
To the best of our knowledge, this is the first work addressing both overhead and straggler issues simultaneously in FL with provable privacy guarantee.


The rest of the paper is organized as follows. First we discuss related works in Section \ref{Sec:Related Work}. Then we present the system model and problem statement in Section \ref{Sec:System Model}. Next, we propose our privacy mechanism in Section \ref{Sec:our scheme}. We theoretically analyze the privacy guarantee and convergence bound, and optimize the noise level in Section \ref{Sec:theoretial}.
We present experimental results in Section \ref{Sec:experiment} and conclude in Section \ref{Sec:Conclusion}.

\section{Related Work}
\label{Sec:Related Work}
\subsection{Secure Multi-Party Computation (SMPC) Schemes}

{\textbf{Pairwise encryption.}}
SMPC based on pairwise encryption (e.g., SecAgg \cite{bonawitz2017practical}) relies on two main principles: 
(1) Clients use pairwise agreements to create masks that hide their parameters while allowing the server to cancel out shared masks. Such schemes impose a significant communication burden due to the large size of encrypted local updates.
(2) In the case of stragglers, clients transmit seeds of their secret sharing to all peers, allowing the server to query for remaining clients and  cancel out masks from stragglers. Such client-to-client transmissions of pairwise random seeds occur {\textit{in each round}}. Otherwise, an adversary could exploit previous reconstruction rounds to infer a client's ground-truth parameters. However, this mask reconstruction period is highly vulnerable to unstable client-to-client links. Even a small probability of failure in each link can lead to a breakdown of the aggregation recovery at the server side, potentially causing training divergence.

\textbf{Homomorphic encryption (HE).}
HE uses a pair of public and private keys to encrypt data while preserving functional features. The threshold Paillier scheme \cite{damgaard2001generalisation,paillier1999efficient} is adopted in FL to enable addition over encrypted parameters, with the shares of the secret key distributed among all clients.
However, a key drawback of HE lies in its significant computation and communication overheads. For instance, in a common cryptosystem with sufficient privacy, each element of the multi-dimensional parameter in FL is encrypted from a 32-bit plaintext to a 4096-bit ciphertext.
The straightforward method to apply HE in FL also relies on a trusted party, which raises privacy concerns on the trusted party. Although improved methods have been proposed to eliminate the need for a trusted third party, they come with additional overheads \cite{hamidi2022efficient}.


{\textbf{Coded computation. }}
Recent SMPC schemes leverage coded computation \cite{jahani2023swiftagg+,so2022lightsecagg} for secure aggregation. The SwiftAgg+ \cite{jahani2023swiftagg+} and LightSecAgg \cite{so2022lightsecagg} protocols use polynomial functions to encode local parameters for client-to-client transmissions. Using the Lagrange interpolation rule, the server can recover the real aggregation as long as a threshold number of clients respond.
However, these schemes heavily rely on client-to-client links. A threshold number of ``non-stragglers"  must succeed in all transmission links to their peers and the server, which is often impractical due to the fragility of client-to-client links. With many failed client-to-client links, the system would fail with training divergence.


All these SMPC schemes ensure secure individual updates from local clients to the server. However, the revealed aggregation at the server side is unperturbed, which will break DP and disclose sensitive information from the dataset. 

\subsection{Differential Privacy (DP)}
DP provides a metric to ensure that a change in a single sample cannot noticeably increase the likelihood of observed outcomes \cite{dwork2014algorithmic}. 
To apply DP in FL, each client can locally add random noise (e.g., Gaussian) to perturb its local parameter \cite{kairouz2021distributed,abadi2016deep, wei_federated_2020,yang2023privatefl}. Despite secure against any untrusted peers or the server, this local noise-adding results in excessive accumulated noise in the aggregation, significantly reducing the model accuracy.

Combining SMPC with DP ensures that individual client-to-server updates are fully secure, requiring only minimal noise for local parameters to satisfy DP. However, such combinations still face limitations inherent to SMPC.
For instance, significant overheads arise when combining HE and DP \cite{truex2019hybrid}. Combining pairwise encryption with large masks and DP lacks straggler resilience \cite{mugunthan2019smpai}. Integrating SecAgg and DP results in high overheads and relies on fragile P2P links \cite{jiang2024dordis}.

\section{System Model and Problem Statement}
\label{Sec:System Model}
In this section, we introduce the system model, then the considered attack model in FL systems, and finally the problem statement of this work.

\subsection{System Model}

We consider an FL system consisting of one central server and $N$ clients in a set $\mathcal{N}$.  
Each client $i \in \mathcal{N}$ holds a local dataset $\mathcal{D}_i=\{\bm{x}_k,y_k\}_{k=1}^{|\mathcal{D}_i|}$, where each data sample contains a feature $\bm{x}_k$ and a label $y_k$, and $| \cdot |$ denotes the cardinality of a set. 
The learning objective is to find the optimal parameter $\bm{\omega}^*$, which minimizes the global 
loss $F(\cdot)$ over the dataset $\mathcal{D}=\cup_{i \in \mathcal{N}}\mathcal{D}_i$ among all clients:
\begin{equation}
\label{Equ: global w*}
\bm{\omega}^*=\arg \min_{\bm{\omega}} 
\left\lbrace 
F(\bm{\omega},\mathcal{D}) 
\triangleq
\sum_{i \in \mathcal{N}} 
\frac{ |\mathcal{D}_i | }{
\sum_{i' \in \mathcal{N} } |\mathcal{D}_{i'} |
}
F_i(\bm{\omega},\mathcal{D}_i)
\right\rbrace,
\end{equation}
where $F_i(\cdot)$ is the local loss function of client $i$.

In FL, clients collaboratively train the global model under the coordination of the server. We consider $T$ training rounds. At the beginning of the FL process, the server first initializes the global parameter as $\bm{\omega}^{(0)}$.  
In a training round $t \in \mathcal{T}=\{1,\cdots,T\}$, the server broadcasts the global model $\bm{\omega}^{(t-1)}$ to all clients.
After receiving $\bm{\omega}^{(t-1)}$, client $i \in \mathcal{N}$ performs local training iteratively over his local dataset $\mathcal{D}_i$ for $E$ epochs and obtain the local parameter $\bm{\omega}_i^{(t)}$. Specifically, in each epoch, client $i$ updates the local parameter by performing the stochastic gradient descent (SGD) steps:
\begin{equation}
\label{Equ:SGD}
\bm{\omega}_i^{(t)}:=\bm{\omega}_i^{(t)}-\eta \nabla F_i(\bm{\omega}_i^{(t)}; \xi_i^{(t)}),
\end{equation}
where the local parameter is initialized as the global parameter $\bm{\omega}^{(t-1)}$, $\eta$ is the learning rate, and $\xi_i^{(t)}$ is the mini-batch sampled from $\mathcal{D}_i$ at round t. 
%
%
%
After $E$ epochs, each client $i \in \mathcal{N}$ masks the local parameter $\bm{\omega}_i^{(t)}$ using a certain privacy mechanism and uploads $\tilde{\bm{\omega}}_i^{(t)}$, a function of $\bm{\omega}_i^{(t)}$, to the server.
Potential mechanisms include vanilla local noise-adding   \cite{wei_federated_2020,abadi2016deep}, coded computing   \cite{jahani2023swiftagg+}, homomorphic encryption  \cite{truex2019hybrid} and our privacy mechanism to be introduced. 

In the process of clients uploading parameters, we address the \textit{straggler issue}, which arises from temporarily disconnected or delayed clients due to weak network connections and inadequate computational resources.
We define this subset of stragglers 
as $\mathcal{N}_{\text{S}}^{(t)} \subset \mathcal{N}$ at round $t$. The exact size and composition of {\color{black}$\mathcal{N}_{\text{S}}^{(t)}$} are generally unknown, as predicting which clients will lag is often impractical. Instead, we assume the maximum number of stragglers as $|\mathcal{N}_{\text{S}}^{(t)}|=S^{(t)} \le \bar{S}$ and $\mathcal{N}_{\text{S}}^{(t)}$ at different rounds are independent with each other.

After receiving 
parameters $\tilde{\bm{\omega}}_i^{(t)}$ from all non-stragglers $i \in \mathcal{N} \setminus \mathcal{N}_{\text{S}}^{(t)}$, the server makes the aggregation to update the global parameter based on a predefined function $\mathcal{A}$ as
\begin{equation}
\bm{\omega}^{(t)}=\mathcal{A}
\left( 
\{\tilde{\bm{\omega}}_i^{(t)}, i \in \mathcal{N} \setminus \mathcal{N}_{\text{S}}^{(t)}\}
\right).
\end{equation}
Using the FedAvg algorithm \cite{mcmahan2017communication}, we have
\begin{equation}
\label{Equ:FedAvg aggregation}
\mathcal{A}
\left( 
\{\tilde{\bm{\omega}}_i^{(t)}, i \in \mathcal{N} \setminus \mathcal{N}_{\text{S}}^{(t)}\}
\right) =
\sum_{i \in \mathcal{N} \setminus \mathcal{N}_{\text{S}}^{(t)}}
\frac{ |\mathcal{D}_i | \tilde{\bm{\omega}}_i^{(t)}}{
\sum_{i' \in \mathcal{N} \setminus \mathcal{N}_{\text{S}}^{(t)}}  |\mathcal{D}_{i'} |
}
\end{equation}
and $\bm{\omega}^{(t)}=\frac{\sum_{i \in \mathcal{N} \setminus \mathcal{N}_{\text{S}}^{(t)}}\tilde{\bm{\omega}}_i^{(t)}}{N-S^{(t)}}$ when all datasets are of the same size.  
Other aggregation rules include  \cite{li2020federated,karimireddy2020scaffold}.  
It is notable that even a single straggler may significantly impact the overall efficiency and convergence of FL, especially when a privacy mechanism requires all clients' participation to recover the aggregation (e.g., \cite{bonawitz2017practical}).
Upon aggregation, the server sends the global parameter $\bm{\omega}^{(t)}$ to all clients for the next round.
After a sufficient number of training rounds, the global parameter $\bm{\omega}^{(T)}$ should eventually converge to the global optimum $
\bm{\omega}^*$ in (\ref{Equ: global w*}).


\subsection{Attack Model}
Although each client $i$ keeps his dataset $\mathcal{D}_i$ locally, the uploaded parameter $\tilde{\bm{\omega}}_i$ and the aggregated global parameter $\bm{\omega}$ might expose sensitive client information under specific attacks such as data reconstruction or model inversion.

In this paper, we examine a scenario with a \textit{semi-honest} (also known as ``honest-but-curious'') setting. All parties honestly follow the training protocol, but an adversary may collude with the untrusted server and a subset of colluding clients in a set $\mathcal{N}_{\text{C}} \subset \mathcal{N}$. It performs \textit{unknown attacks} to infer other clients' sensitive information. 
We assume the maximum size of the colluder set as $|\mathcal{N}_{\text{C}}|=C\le \bar{C}$. 
The exact size and composition of $\mathcal{N}_{\text{C}}$ are not known. Hence, the subsets $\mathcal{N}_{\text{S}}^{(t)}$ and $\mathcal{N}_{\text{C}}$ may either intersect or distinct.

Given this setting, our goal is to design a privacy mechanism under the worst scenario, characterized by the unknown attacks and an unknown colluder set.
Unlike the standard SMPC alone mechanisms \cite{jahani2023swiftagg+,bonawitz2017practical}, we consider an \textit{untrusted server}. 
We aim to protect all potentially exposed parameters, including individual client parameters $\tilde{\bm{\omega}}_i$ and the global aggregation $\bm{\omega}$, to provide a robust and strong privacy guarantee. This motivates us to use differential privacy as our privacy metric.

\subsection{Problem Statement}

To address the privacy concerns from unknown attacks, we leverage the concept of Differential Privacy (DP) introduced in   \cite{dwork2014algorithmic}. 
The formal definition is presented as follows:

\begin{defn}[Differential Privacy]
	\label{Defn:DP}
	A randomized algorithm $\mathcal{P}: \mathcal{X} \rightarrow \mathcal{R}$  with domain $\mathcal{X}$ and range $\mathcal{R}$ is $(\varepsilon,\delta)$-differentially-private if, for all $\mathcal{S} \subseteq \mathcal{R}$ and for all $d,d' \in \mathcal{X}$ such that $||d-d'||_1 \leq 1$:
	\begin{equation*}
		\operatorname{Pr}[\mathcal{P}(d) \in \mathcal{S}] \leq e^\varepsilon \operatorname{Pr}[\mathcal{P}(d') \in \mathcal{S}]+\delta,
	\end{equation*}
	where $\varepsilon$ denotes the distinguishability bound of all outputs on adjacent inputs $d$ and $d'$. $\delta$ represents the event that the ratio of the probabilities for two adjacent inputs $d$ and $d'$ cannot be bounded by $e^\varepsilon$. 
\end{defn}

In our system model, we carefully consider all potentially exposed parameters, including the local updates $\tilde{\bm{\omega}}_j^{(t)}, \forall j \in \mathcal{N}\setminus \mathcal{N}_{\text{S}}^{(t)}$ and the global aggregation $\bm{\omega}^{(t)}$ in each round $t \in \mathcal{T}$. To quantify the privacy loss for a specific client $i$ within the FL system, we introduce the privacy loss function for a privacy mechanism $\mathcal{M}$ as in  \cite{abadi2016deep}:
\begin{equation}
\label{Equ: privacy loss def}
\mathcal{L}_i(\mathcal{M}) = \ln \frac{\operatorname{Pr}(\mathcal{M}(\mathcal{D}))}{\operatorname{Pr}(\mathcal{M}(\mathcal{D}'))},
\end{equation}
where $\mathcal{M}(\mathcal{D})=\{\tilde{\bm{\omega}}_j^{(t)}, j \in \mathcal{N}\setminus \mathcal{N}_{\text{S}}^{(t)}; {\bm{\omega}^{(t)}},  t \in \mathcal{T}\}$ represents all outcomes of mechanism $\mathcal{M}$ based on dataset $\mathcal{D}$ during all training rounds. $\mathcal{D}$ and $\mathcal{D}'$ are two adjacent datasets where only one data sample in client $i$'s datasets $\mathcal{D}_i$ and $\mathcal{D}_i'$ differs. 
Following Definition \ref{Defn:DP}, the $(\varepsilon,\delta)$-DP guarantee is equivalent to
\begin{equation}
\label{Inequ:privacy loss requirement}
	\operatorname{Pr}(|\mathcal{L}_i(\mathcal{M})| \geq \varepsilon) \leq \delta.
\end{equation}

The rigorous privacy constraints stated in (\ref{Inequ:privacy loss requirement}) is the strongest DP against an untrusted server, various attacks, and unknown colluder and straggler sets. In addition to satisfying this, the designed mechanism should also be lightweight with low overheads, straggler-resilient in all client-to-server and client-to-client links, and approach the best possible accuracy performance.
To the best of our knowledge, none of the existing schemes can achieve these goals simultaneously. 

\section{The Proposed LightDP-FL Scheme}
\label{Sec:our scheme}

In this section, we introduce our privacy mechanism combining pairwise noise terms with individual noise terms in FL systems.
We then discuss how the combination 
disturbs local and global parameters, which is important for further privacy analysis and noise design with straggler concerns. 

%


\subsection{Generation of Pairwise Noise}
We propose the simple idea of adding common randomness additionally between each pair of clients to the local parameters, i.e., $\frac{N(N-1)}{2}$ pairwise noise terms are generated in total. Specifically, at the individual client's end, each one of
the pairwise noise terms alone helps to provide necessary randomness for DP. At the server side, most pairwise noise terms can be effectively canceled, leading to improved accuracy performance.
Even in the presence of stragglers, the pairwise noise can  be designed small enough in Section \ref{Sec:theoretial}.

To establish common randomness and generate pairwise shared noise in each round, each pair of clients 
$i,j\in \mathcal{N}$ agree on a common seed $s$ and a pseudorandom generator (PRG) based on the Diffie-Hellman key agreement protocol   \cite{li2010research}. This comes with a single communication round before the learning process. The PRG should generate the pairwise noise term $\bm{r}_{ij}^{(t)}=\bm{r}_{ji}^{(t)}$ following a Gaussian distribution, with mean 0 and variance $\sigma_{ij}^2=\sigma_{ji}^2$, based on the common seed and the training round $t$  \cite{bonawitz2017practical}.
%
For simplicity, we use the notation $\bm{r}_{ij}$ and omit the subscript $(t)$ for the remainder of this paper, focusing on a single training round. This notation simplification applies to other parameters as well, when there is no ambiguity.

\subsection{Local Parameter Updates}
The pairwise noise terms alone, however, are insufficient to attain DP. This is because, in the absence of stragglers, the untrusted server can cancel out all pairwise noise terms, thereby revealing the exact sum without any perturbation.
Hence, we further introduce an individual Gaussian noise term $\bm{n}_i$ together with $N-1$ pairwise noise terms $\bm{r}_{ij}$'s for each client $i \in \mathcal{N}$. Specifically, client $i$ masks the local update $\bm{\omega}_i$ by adding individual  noise $\bm{n}_i \sim \mathcal{N}(0,\sigma_i^2 \mathbf{I})$ and pairwise noise terms $\bm{r}_{ia} \sim \mathcal{N}(0,\sigma_{ia}^2 \mathbf{I})$ for all $a > i, a\in \mathcal{N}$, and subtracting $\bm{r}_{bi} \sim \mathcal{N}(0,\sigma_{bi}^2 \mathbf{I})$ for all $b < i, b\in \mathcal{N}$, as follows:
\begin{equation}
\label{Equ:sum of pairwise+DP}
\tilde{\bm{\omega}}_i:=\bm{\omega}_i+\sum_{a\in \mathcal{N}:a>i} \bm{r}_{ia}-\sum_{b\in \mathcal{N}:b<i} \bm{r}_{bi}
+\bm{n}_i.
\end{equation}
Notice that here the noise variances $\sigma_i^2$ and $\sigma_{ij}^2, \forall i, j\in \mathcal{N}$ are independent and will be further designed in Section \ref{Sec:theoretial} to satisfy the DP requirement in (\ref{Inequ:privacy loss requirement}). $\mathbf{I}$ is the $d \times d$ identity matrix, where $d$ is the dimension of the model parameters.

Given that colluding clients $j \in \mathcal{N}_{\text{C}}$  reveal the pairwise noise terms $\bm{r}_{ij}$'s for potential attacks, we use a term $\bm{m}_i$ to denote the actual local disturbance for client $i$, i.e.,
\begin{equation}
\label{Equ:local disturbance term m i}
\begin{aligned}
\bm{m}_i&=
\sum_{a:a>i, a \not\in \mathcal{N}_{\text{C}}} \bm{r}_{ia}-\sum_{b:b<i,b\not\in \mathcal{N}_{\text{C}}} \bm{r}_{bi}
+\bm{n}_i \\
&\sim
\mathcal{N}\left( 0,\sum_{j \neq i, j\in \mathcal{N}\setminus\mathcal{N}_{\text{C}}} \sigma_{ij}^2 \mathbf{I}+\sigma_i^2 \mathbf{I}
\right) ,
\end{aligned}
\end{equation}
including his own individual noise 
 and $N-C-1$ unrevealed pairwise terms.


\subsection{Global Aggregation}
After receiving all local updates $\tilde{\bm{\omega}}_i^{(t)}$ from client $i \in \mathcal{N}\setminus \mathcal{N}_{\text{S}}^{(t)}$, the  server  updates the global parameter. For simplicity, we consider the average aggregation  \cite{mcmahan2017communication} based on (\ref{Equ:FedAvg aggregation}):
\begin{equation*}
\label{Equ:aggregated sum}
	\begin{aligned}
{\bm{\omega}}^{(t)} 
=\frac{1}{N-S^{(t)}}\sum_{i \in \mathcal{N}\setminus \mathcal{N}_{\text{S}}^{(t)}}
\bm{\omega}_i^{(t)} + \bm{n}_D^{(t)},
	\end{aligned}
\end{equation*}
where 
%
the accumulated noise term includes the uncanceled pairwise and all individual noise terms:
\begin{equation}
\label{Equ:n_D}
\begin{aligned}
&\bm{n}_D^{(t)}\\
=
&\frac{\sum_{i \in \mathcal{N} \setminus \mathcal{N}_{\text{S}}^{(t)}} 
		\left(
		\sum_{a\in \mathcal{N}_{\text{S}}^{(t)}:a>i} \bm{r}_{ia}^{(t)}-\sum_{b\in \mathcal{N}_{\text{S}}^{(t)}:b<i} \bm{r}_{bi}^{(t)}
		+\bm{n}_i^{(t)}
		\right) }{N-S^{(t)}}
		\\
\sim 
&\mathcal{N}\left( 0,
\frac{\sum_{i \in \mathcal{N} \setminus \mathcal{N}_{\text{S}}^{(t)}}
\left( 
\sum_{j\in \mathcal{N}_{\text{S}}^{(t)}} \sigma_{ij}^2 
\mathbf{I}+\sigma_i^2 \mathbf{I}
\right)}{(N-S^{(t)})^2}
\right) . 
\end{aligned}
\end{equation}

\begin{figure}[t]
\centering
\centerline{\includegraphics[width=2.2 in]{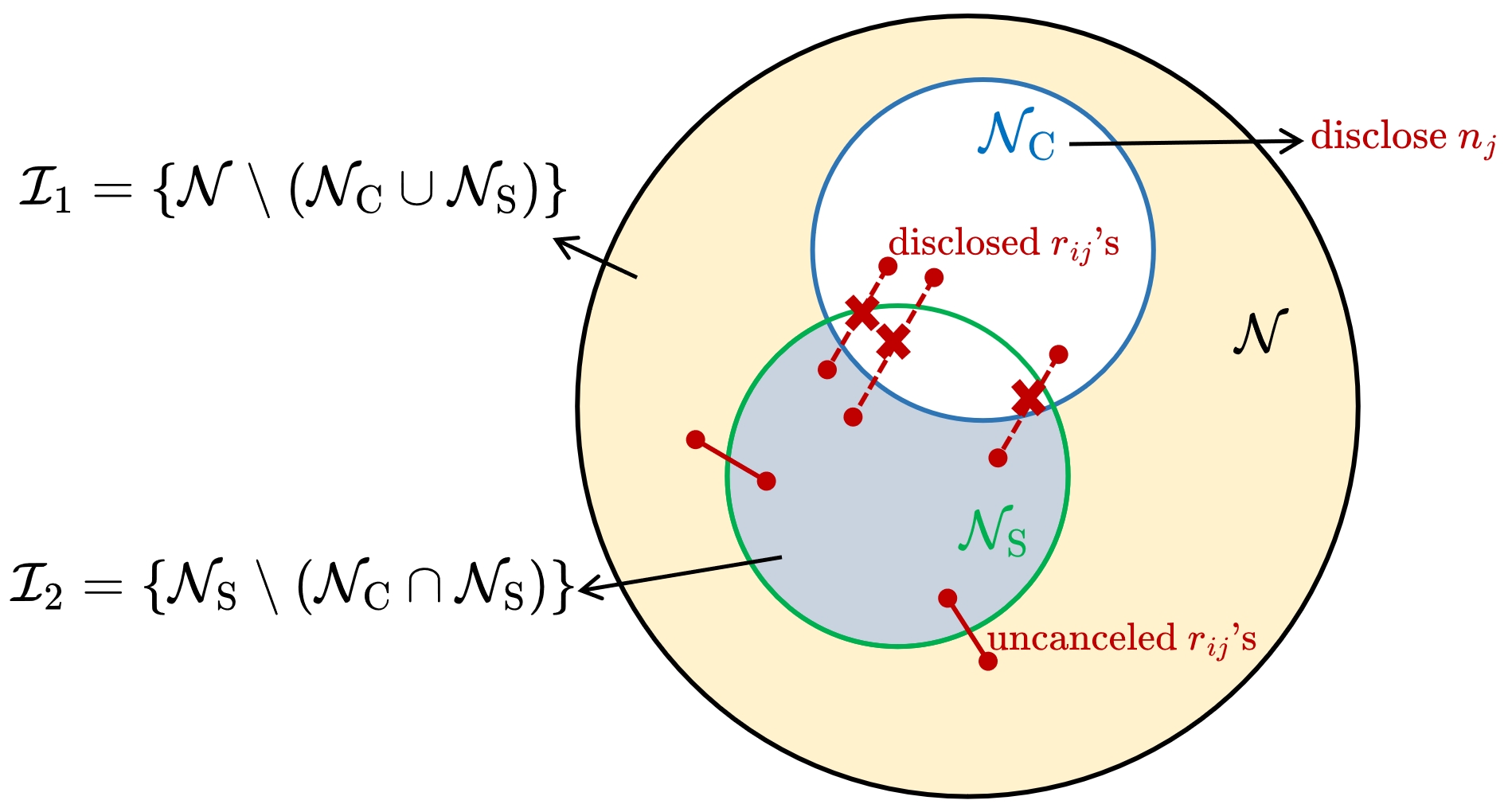}}
\caption{
Illustration of summed noise terms $\bm{n}_D$ in (\ref{Equ:n_D}) and $\bm{m}_D$ in (\ref{Equ:m_D}).
1) Due to the straggler issue, pairwise noise terms $\bm{r}_{ij}, \forall i \in \mathcal{N} \setminus \mathcal{N}_{\text{S}},j \in \mathcal{N}_{\text{S}}$ remain uncanceled in the aggregated noise term $\bm{n}_D$ in (\ref{Equ:n_D}),  represented by red lines. 
2) Remaining colluders $j \in \mathcal{N}_{\text{C}}\setminus(\mathcal{N}_{\text{C}} \cap \mathcal{N}_{\text{S}})$ will disclose their individual terms $\bm{n}_j$ and pairwise terms $\bm{r}_{ij}$'s.
Hence, only $\bm{r}_{ij}$'s with $i \in \mathcal{I}_1=\{\mathcal{N}\setminus (\mathcal{N}_{\text{C}} \cup \mathcal{N}_{\text{S}}) \}$ and $j \in \mathcal{I}_2= \{\mathcal{N}_{\text{S}}\setminus (\mathcal{N}_{\text{C}} \cap \mathcal{N}_{\text{S}}) \}$, and noise terms $\bm{n}_j$'s with $j \in \mathcal{I}_1$ remain in the global disturbance term $\bm{m}_D$ for privacy.
}
\label{Fig:colluder keys}
\end{figure}

Moreover, the remaining pairwise $\bm{r}_{ij}^{(t)}$'s and individual noise terms $\bm{n}_j^{(t)}$'s in $\bm{n}_D^{(t)}$ from (\ref{Equ:n_D}) might be disclosed by colluding clients $j \in \mathcal{N}_{\text{C}}$. As illustrated in Fig.~\ref{Fig:colluder keys} (subscript omitted for clarity), only $\bm{r}_{ij}^{(t)}$'s for  all $i \in \mathcal{I}_1^{(t)}=\{\mathcal{N}\setminus (\mathcal{N}_{\text{C}} \cup \mathcal{N}_{\text{S}}^{(t)}) \}$, $j \in \mathcal{I}_2^{(t)}= \{\mathcal{N}_{\text{S}}^{(t)}\setminus (\mathcal{N}_{\text{C}} \cap \mathcal{N}_{\text{S}}^{(t)}) \}$, and
$\bm{n}_i$'s for all $i \in \mathcal{I}_1^{(t)}$ 	remain in the global disturbance term to provide privacy guarantee, denoted by  $\bm{m}_D^{(t)}$:
\begin{equation}
\label{Equ:m_D}
\begin{aligned}
\bm{m}_D^{(t)} 
&=
\frac{\sum_{i \in \mathcal{I}_1^{(t)}} 
\left(
\sum_{a\in \mathcal{I}_2^{(t)}:a>i} \bm{r}_{ia}^{(t)} 
-\sum_{b\in \mathcal{I}_2^{(t)}:b<i} \bm{r}_{bi}^{(t)}
+\bm{n}_i
\right)}{N-S}
 \\
&\sim 
\mathcal{N}
\left( 
0, \frac{\sum_{i \in \mathcal{I}_1^{(t)},j \in \mathcal{I}_2^{(t)}} \left( \sigma_{ij}^2 \mathbf{I} +\sigma_i^2 \mathbf{I} \right) }{(N-S^{(t)})^2} 
\right) .
\end{aligned}
\end{equation}
%


\section{Theoretical Analysis
}
\label{Sec:theoretial}
In this section, we establish the privacy constraints and rigorously prove the DP constraint,  while accounting for all possible outcomes. 
Subsequently, we analyze the convergence bound for the FL system. We then formulate an optimization problem for convergence bound minimization and design the noise variances under the DP guarantee.


\subsection{Privacy Analysis}
As the pairwise noise terms $\bm{r}_{ij}$'s among clients are coupled, the joint outcomes of all local uploads and global aggregations are correlated, making the DP analysis not straightforward. However, our results prove the sufficient conditions satisfying (\ref{Inequ:privacy loss requirement}) to achieve the $(\varepsilon,\delta)$-DP for each client $i$, by considering the joint normal distribution of all outcomes. 

Recall that colluders in set $\mathcal{N}_{\text{C}}$ reveal all pairwise and individual noise terms they hold, while stragglers in set $\mathcal{N}_{\text{S}}$ fail to upload the parameters in time. Hence, we only care about the privacy for clients in set $\mathcal{I}_1=\mathcal{N}\setminus\left( \mathcal{N}_{\text{C}} \cup \mathcal{N}_{\text{S}}\right) $.
We also use $\mathcal{I}_2=\mathcal{N}_{\text{S}}\setminus\left( \mathcal{N}_{\text{C}} \cap \mathcal{N}_{\text{S}}\right) $ to denote the set of stragglers but not colluders (i.e., who drop out leaving the pairwise noise terms uncanceled and unrevealed).
Then all possible outcomes that an adversary can observe include the two parts: individual updates $\tilde{\bm{\omega}}_i$ from client $i \in \mathcal{I}_1$ and global aggregation $\bm{\omega}$.

For client $i \in \mathcal{I}_1$'s individual update $\tilde{\bm{\omega}}_i$, let $\mathcal{M}_i(\mathcal{D}_i) = \bm{\omega}_i(\mathcal{D}_i) + \bm{m}_i$ represent the local outcome with dataset $\mathcal{D}_i$, excluding those noise terms potentially revealed by colluders.
$\bm{\omega}_i(\mathcal{D}_i)$ is the parameter to be protected, and $\bm{m}_i$ in (\ref{Equ:local disturbance term m i}) provides all the random disturbance to ensure privacy.
For the global aggregation $\bm{\omega}$, let $\mathcal{M}_D(\mathcal{D})=\sum_{j \in \mathcal{I}_1} \bm{\omega}_j(\mathcal{D}_j)+\bm{m}_D$ represent the global privacy-ensured outcome for the sensitive aggregation $\sum_{j \in \mathcal{I}_1} \bm{\omega}_j(\mathcal{D}_j)$, where $\mathcal{D}=\cup_{j \in \mathcal{I}_1} \mathcal{D}_j$ is all privacy-sensitive clients' datasets. The global disturbance term $\bm{m}_D$ is given in (\ref{Equ:m_D}), excluding all noise terms revealed by colluders.

We consider the privacy loss random variable $\mathcal{L}_i$ in (\ref{Equ: privacy loss def}) for a single client $i \in \mathcal{I}_1$ based on his adjacent datasets $\mathcal{D}_i$ and $\mathcal{D}_i'$:
\begin{equation}
\label{Equ: privacy loss-initial ratio form}
\begin{aligned}
&\mathcal{L}_i\\
=&\ln 
\begin{array}{c}
\frac
{
\operatorname{Pr}\left(
\begin{array}{l}
\mathcal{M}_i(\mathcal{D}_i)=\bm{\omega}_i(\mathcal{D}_i)+\bm{x}_i;\\
\forall j \in \mathcal{I}_1, j\neq i:
\mathcal{M}_j\left(\mathcal{D}_j\right)=\bm{\omega}_j\left(\mathcal{D}_j\right)+\bm{x}_j ; \\
\mathcal{M}_D(\mathcal{D})=\bm{\omega}_i(\mathcal{D}_i)+\sum_{j \neq i, j\in \mathcal{I}_1}\bm{\omega}_j(\mathcal{D}_j) +\bm{x}_D
\end{array}
\right) 
}
{
\operatorname{Pr}\left(
\begin{array}{l}
\mathcal{M}_i(\mathcal{D}_i')=\bm{\omega}_i(\mathcal{D}_i)+\bm{x}_i;\\
\forall j \in \mathcal{I}_1, j\neq i: 
\mathcal{M}_j\left(\mathcal{D}_j\right)=\bm{\omega}_j\left(\mathcal{D}_j\right)+\bm{x}_j ; \\
\mathcal{M}_D(\mathcal{D}')=\bm{\omega}_i(\mathcal{D}_i)+\sum_{j \neq i, j\in \mathcal{I}_1}\bm{\omega}_j(\mathcal{D}_j) +\bm{x}_D
\end{array}
\right) 
}
\end{array} \\
=
&\ln 
\begin{array}{c}
\frac
{
\operatorname{Pr}\left(
\begin{array}{l}
\mathcal{M}_i(\mathcal{D}_i)=\bm{\omega}_i(\mathcal{D}_i)+\bm{x}_i;\\
\forall j \in \mathcal{I}_1, j\neq i: 
\mathcal{M}_j\left(\mathcal{D}_j\right)=\bm{\omega}_j\left(\mathcal{D}_j\right)+\bm{x}_j 
\end{array}
\right) 
}
{
\operatorname{Pr}\left(
\begin{array}{l}
\mathcal{M}_i(\mathcal{D}_i')=\bm{\omega}_i(\mathcal{D}_i')+\bm{v}+\bm{x}_i;\\
\forall j \in \mathcal{I}_1, j\neq i: 
\mathcal{M}_j\left(\mathcal{D}_j\right)=\bm{\omega}_j\left(\mathcal{D}_j\right)+\bm{x}_j 
\end{array}
\right) 
}
\end{array}
\\
=
&\ln
\begin{array}{c}
\frac
{
\operatorname{Pr}\left(
\begin{array}{l}
\bm{m}_i=\bm{x}_i;\\
\forall j \in \mathcal{I}_1, j\neq i: 
\bm{m}_j=\bm{x}_j
\end{array}
\right) 
}
{
\operatorname{Pr}\left(
\begin{array}{l}
\bm{m}_i=\bm{x}_i+\bm{v};\\
\forall j \in \mathcal{I}_1, j\neq i: 
\bm{m}_j=\bm{x}_j
\end{array}
\right) 
}
\end{array}
\end{aligned}
\end{equation}
for a specific time slot\footnote{
In this section, we  consider the privacy loss for a single time round to establish the constraints of the appropriate noise variance sufficient for  a DP guarantee. 
Given that the model parameters are disclosed $T$ times throughout the learning process, we can simply apply the composition theorem of DP to address the cumulative privacy loss across multiple rounds \cite{dwork2014algorithmic}. According to this theorem, the cumulative privacy risk over $T$ rounds can be effectively managed by scaling all noise variances by a factor of $\sqrt{T}$. 
} with $C$ colluders and $S$ stragglers, 
where $\bm{v}=\bm{\omega}_i(\mathcal{D}_i)-\bm{\omega}_i(\mathcal{D}_i')$. 
Note that we omit the global disturbance term $\bm{m}_D$ in the second equality as it is linearly dependent on $\bm{m}_j,\forall j \in \mathcal{I}_1$ under the average aggregation\footnote{Our mechanism can be adapted to other aggregation rules where $\bm{m}_D$ might be kept in the privacy loss function.}. In other words, we can prove that the determinant of the covariance matrix among $(\bm{m}_i,\forall i \in \mathcal{I}_1;\bm{m}_D)$ is 0. 

Since we need to calculate the probabilities in (\ref{Equ: privacy loss-initial ratio form}), we first study the joint density function for all $\bm{m}_j, j\in \mathcal{I}_1$. Then, we derive the mean and variance of $\mathcal{L}_i$. This will then allow us to study the DP guarantee.

\subsubsection{Joint distribution of all disturbance terms}	
First we are interested to give the joint density function of all disturbance terms, i.e., $f(\bm{m}_j,\forall j \in \mathcal{I}_1)$. 
With a bit of abusing notations, we use $\{1,\cdots,|\mathcal{I}_1|\}$ to denote the set of clients in set $\mathcal{I}_1$, and omit the identity matrix $\mathbf{I}$ in the noise variances. 

\begin{lemma}
\label{Lemma:covariance matrix}
Let 
\begin{equation}
\label{Matrix:C_m}
C_{\bm{m}}=
\bordermatrix{
& \bm{m}_1 &\cdots &\bm{m}_{|\mathcal{I}_1|} 
\cr
\bm{m}_1
&C_{11} & \cdots &C_{1 |\mathcal{I}_1|} 
\cr
\ \vdots	& \vdots & \ddots & \vdots  
\cr
\bm{m}_{|\mathcal{I}_1|}		
&C_{|\mathcal{I}_1|1} &\cdots &C_{|\mathcal{I}_1||\mathcal{I}_1|} 
}
\end{equation}
be an $|\mathcal{I}_1| \times |\mathcal{I}_1|$ covariance matrix with
\begin{equation*}
C_{ii}
=\sum_{j \neq i, j\in \mathcal{N}\setminus\mathcal{N}_{\text{C}}} \sigma_{ij}^2+\sigma_i^2
=\sum_{j \neq i, j\in \mathcal{I}_1+\mathcal{I}_2} \sigma_{ij}^2+\sigma_i^2
\end{equation*}
for $i \in \mathcal{I}_1$ and
\begin{equation*}
C_{ij}=C_{ji}
=-\sigma_{ij}^2
\end{equation*}
for $i,j \in \mathcal{I}_1$, $i \neq j$,
then the multivariate Gaussian distribution for $(\bm{m}_1, \cdots, \bm{m}_{|\mathcal{I}_1|})$ is  
\begin{equation}
\label{Equ:joint distribution for all obs m}
		\begin{aligned}
&f_{\bm{m}_1, \cdots, \bm{m}_{|\mathcal{I}_1|}}
\left(\bm{x}_1, \cdots, \bm{x}_{|\mathcal{I}_1|};
C_{\bm{m}}
\right)   \\
			&=
\frac{\left( det\left( \mathbf{C}_{\bm{m}}^{-1}\right)\right)  ^{1 / 2}}{(2 \pi)^{|\mathcal{I}_1| / 2}} \exp \left\{-\frac{\mathbf{X}^t \mathbf{C}_{\bm{m}}^{-1}\mathbf{X}}{2}\right\},
		\end{aligned}
	\end{equation}
where $det(\cdot)$ is the determinant of a matrix, $C_{\bm{m}}^{-1}$ is the inverse of 
$C_{\bm{m}}$, $\mathbf{X}=[\bm{x}_1, \cdots, \bm{x}_{|\mathcal{I}_1|}]$, and $\mathbf{X}^t$ is its transpose matrix.	
	
\end{lemma}


%

\subsubsection{Characterizing the mean and variance of $\mathcal{L}_i$}
Next, we take the multivariate distribution of $\bm{m}$ in (\ref{Equ:joint distribution for all obs m}) into the private loss function in (\ref{Equ: privacy loss-initial ratio form}) to rewrite the random variable $\mathcal{L}_i$.
	
\begin{lemma}
\label{Lemma:write privacy loss as Gaussian}
Let $\mathcal{D}_i$ and $\mathcal{D}_i'$ be two adjacent datasets with only one different sample from client $i \in \mathcal{I}_1$, then the privacy loss random variable $\mathcal{L}_i$ for client $i$ under our mechanism 
follows a Gaussian distribution with mean $\frac{C_{i i}^{-1}}{2} \|\bm{\omega}_i(\mathcal{D}_i)-\bm{\omega}_i(\mathcal{D}_i')\|_2^2$ and variance
$\left[
\sum_{j \in \mathcal{I}_1} (C_{i j}^{-1})^2
C_{j j}    
\right] 
\|\bm{\omega}_i(\mathcal{D}_i)-\bm{\omega}_i(\mathcal{D}_i')\|_2^2$,
%
%
where $\bm{\omega}_i(\mathcal{D}_i)$ is client $i$'s local parameter with dataset $\mathcal{D}_i$.
	\end{lemma}
%

\subsubsection{Sufficient Conditions for DP Guarantee}
To prove the $(\varepsilon,\delta)$-DP guarantee for each client $i$, we need to ensure the absolute value of the privacy loss random variable exceeds $\varepsilon$ with probability at most $\delta$, i.e., (\ref{Inequ:privacy loss requirement}). 

\begin{prop}
\label{Prop:general DP}  
To satisfy $(\varepsilon,\delta)$-DP privacy for all clients in $\mathcal{I}_1$, the following inequality set gives the sufficient condition: 
\begin{equation}
\label{Inequ:general DP constraint}
\frac{1}{
\sqrt{\sum_{j \in \mathcal{I}_1} (C_{i j}^{-1})^2
		C_{j j}   }
}
\geq 
\frac{\sqrt{2\log(2/\delta)}\Delta}{\varepsilon}, 
\forall i \in \mathcal{I}_1, \forall \mathcal{I}_1, \mathcal{I}_2,
\end{equation}
where $\Delta=\max_{\mathcal{D}_i,\mathcal{D}_i':d(\mathcal{D}_i,\mathcal{D}_i')=1} ||\bm{\omega}_i(\mathcal{D}_i)-\bm{\omega}_i(\mathcal{D}_i')||_2$ is the sensitivity of local parameter based on two adjacent datasets.
\end{prop}
\begin{IEEEproof}
See Appendix \ref{Proof:Prop:general DP}.
\end{IEEEproof}

Proposition \ref{Prop:general DP} gives the feasible set for all noise variances within the DP guarantee.
Given that the exact sizes and compositions of subsets $\mathcal{N}_{\text{C}}$ and $\mathcal{N}_{\text{S}}$ are unknown, the examination of all potential combinations of the sets $\mathcal{I}_1$ and $\mathcal{I}_2$ already considers the worst case to provide the robust privacy guarantee.

Next, we will analyze the accuracy performance in the learning system for further design of the noise variances under the privacy constraints.

\subsection{Convergence Analysis}
In this section, we first make certain standard assumptions about the global loss functions $F(\cdot)$ and client $i$'s local loss function $F_i(\cdot)$. Then we give the convergence bound for $
\mathbb{E} \{F({\bm{\omega}}^{(T)})-F({\bm{\omega}}^*)\}$ to measure the training accuracy.
\begin{assmp}
	The global loss function $F(\cdot)$ and all client $i$'s local loss functions $F_i(\cdot)$ are convex. 
\end{assmp}

\begin{assmp}
$F_i(\cdot)$ is $\beta$-Lipschitz with
$\left\|F_i(\bm{\omega})-F_i\left(\bm{\omega}^{\prime}\right)\right\| \leq \beta \| \bm{\omega}-\bm{\omega}' \|$, for any $\bm{\omega}$ and $\bm{\omega}'$.
\end{assmp}

\begin{assmp}
$F_i(\cdot)$ is convex, $\rho$-Lipschitz smooth with
$$\left\|\nabla F_i(\bm{\omega})-\nabla F_i\left(\bm{\omega}^{\prime}\right)\right\| \leq \rho \| \bm{\omega}-\bm{\omega}' \|,$$ for any $\bm{\omega}$ and $\bm{\omega}'$.
\end{assmp}

\begin{assmp}
	$F_i(\bm{\omega})$ 
	satisfies the Polyak-Lojasiewicz condition with the positive parameter $l$, which implies that $F(\bm{\omega})-F\left(\bm{\omega}^*\right) \leq \frac{1}{2 l}\|\nabla F(\bm{\omega})\|^2$.
\end{assmp}

\begin{assmp}
	$\mathbb{E}\left\{\left\|\nabla F_i(\bm{\omega})\right\|^2\right\} \leq\|\nabla F(\bm{\omega})\|^2 B^2, \forall i \in \mathcal{N}$.
\end{assmp}

\begin{prop}
\label{Prop:convergence bound}
Conditional on the number of stragglers for all rounds $\mathbf{S}=\{S^{(t)},t \in \mathcal{T}\}$,
the convergence bound of our LightDP-FL mechanism after $T$ rounds is given by
\begin{equation}
\label{Equ:convergence bound}
\begin{aligned}
&\mathbb{E}\left\{F\left({\bm{\omega}}^{(T)}\right)-F\left(\bm{\omega}^*\right)|\mathbf{S}\right\}
		\\ 
\leq
&\left(1+2 l \lambda_2\right)^T \left[F\left(\bm{\omega}^{(0)}\right)-F\left(\bm{\omega}^*\right)\right] \\  
+&
\sum_{t=1}^T
\left(\lambda_1 \beta \mathbb{E}\{\|\mathbf{n}_D^{(t)}\|\}+\lambda_0 \mathbb{E}\left\{\|\mathbf{n}_D^{(t)}\|^2\right\}\right)^t 
\left(1+2 l \lambda_2\right)^{T-t},
	\end{aligned}
\end{equation}
	where $\lambda_0=\frac{\rho}{2}$, $\lambda_1=1+{\rho B}$, $\lambda_2=-1+{\rho B}+\frac{\rho B^2}{2 }$. The global noise $\mathbf{n}_D^{(t)}$ is a function of $S^{(t)}$, as in (\ref{Equ:n_D}).
\end{prop}

We can observe that the term $1+2 l \lambda_2$ in the upper bound is positive. Hence, minimizing the convergence bound in (\ref{Equ:convergence bound}) is equivalent to minimize $\sum_{t=1}^T
\left(\lambda_1 \beta \mathbb{E}\{\|\mathbf{n}_D^{(t)}\|\}+\lambda_0 \mathbb{E}\left\{\|\mathbf{n}_D^{(t)}\|^2\right\}\right)^t $.
{\color{black}
Notice that $\mathbf{n}_D^{(t)}$ is only influenced by the realization of the straggler set $\mathcal{N}_{\text{S}}^{(t)}$ and does not depend on the colluder set because we have ensured the worst-case privacy guarantee for all possible realizations of $\mathcal{N}_{\text{C}}$ in Proposition \ref{Prop:general DP}. 
The actual performance indicated by the bound in (\ref{Equ:convergence bound}) is random, depending on the realized straggler numbers $\mathbf{S}$.
To maintain the expected training performance, later on we will consider the distribution of $S^{(t)} \in [0, \bar{S}]$. 
Such a distribution for $S^{(t)}$ can be obtained, for example, by modeling the successful transmission for each link as a Bernoulli process \cite{lee2021adaptive,gu2021fast}.

}

\subsection{Noise Variance Optimization}
\label{Subsec:noise opt}
Recall that our objective is to minimize the convergence bound under the DP guarantee. 
{\color{black}
The DP constraints in (\ref{Inequ:general DP constraint}) already ensures the worst-case guarantee for any number of stragglers $S^{(t)} \in [0,\bar{S}]$.
Upon that, we aim to optimize the convergence bound by incorporating the distribution of $S^{(t)}$.
We assume that $S^{(t)}$ for any round $t$ follows an i.i.d. distribution $g(s)$, with an upper bound $\bar{S}$. We let $\mathbb{E}\|\mathbf{n}_D^{(t)}\|=\mathbb{E}\|\mathbf{n}_D\|$ and $S^{(t)}=S$ for all $t\in \mathcal{T}$. Thus, minimizing the expected convergence bound in (\ref{Equ:convergence bound}) across all realizations of $\mathbf{S}$ is equivalent to minimizing $\mathbb{E}_{S \sim g(s)} \mathbb{E}\|\mathbf{n}_D(S) \|$ with $\mathbf{n}_D$ in (\ref{Equ:n_D}).


It is important to note that existing works combining SMPC and DP have to rely on the upper bound $\bar{S}$ for the number of stragglers to add DP noise.
This scheme can result in excessive accumulated noise at the server side in case of fewer stragglers than $\bar{S}$. In contrast, LightDP-FL combines individual and pairwise noise terms, which allows for more flexible control of noise variances. We utilize the range $S \in [0,\bar{S}]$ to ensure DP privacy in the worst case, while leveraging any specific distribution for $S$ to derive the expected convergence performance. This is achieved by reformulating the objective function $\mathbb{E}_{S \sim g(s)} \mathbb{E}\|\mathbf{n}_D(S) \|$ for any arbitrary distribution $g(s)$. To derive a clean result, we reformulate the problem as follows.


}

\begin{figure}[!t]
	\centering
	\includegraphics[width=3.5 in]{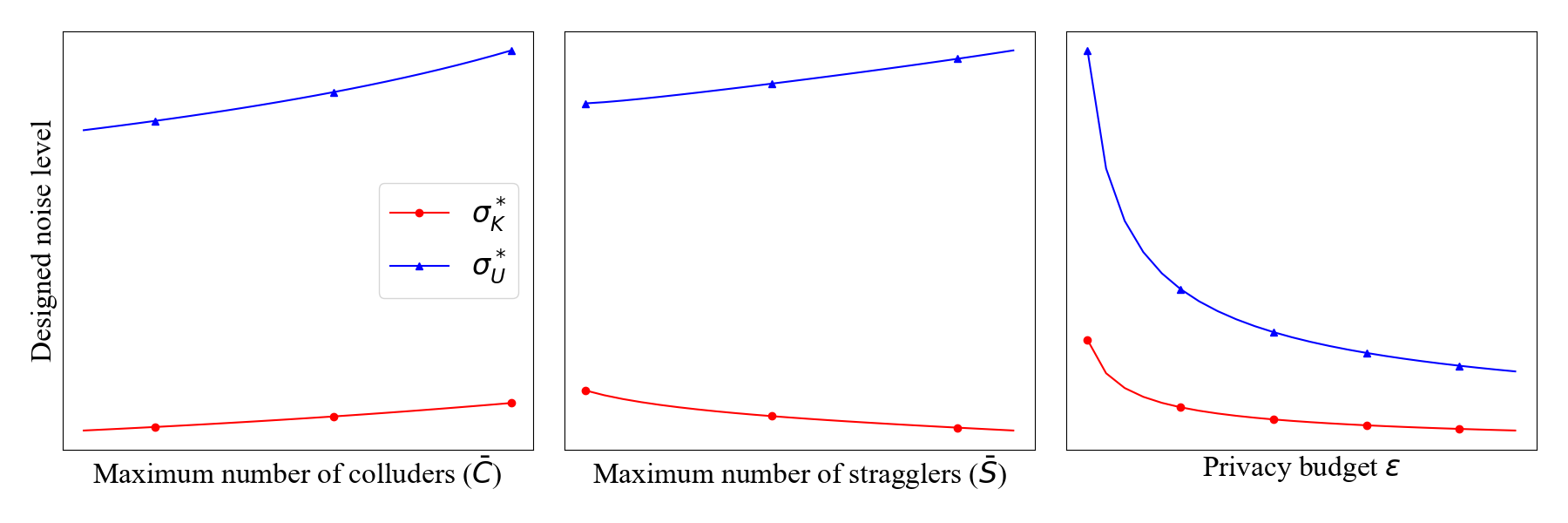}
\caption{Illustration of derived noise levels $(\sigma_\text{K},\sigma_{\text{U}})$ in Example \ref{ex:variance sol-homo}.}
\label{fig:theorem results}
\end{figure}


\begin{prop}
\label{Prop:rewrite opt}
Let $\sigma_{ij}^2=\sigma_{\text{K}}^2, \forall i,j\in \mathcal{N}$ and $\sigma_i^2=\sigma_{\text{U}}^2,\forall i\in \mathcal{N}$ denote the variances for all pairwise noise terms $\bm{r}_{ij}$'s and individual noise terms $\bm{n}_i$'s, respectively. 
Then we write the optimization problem  
equivalently as:
\begin{multline}
\label{Opt:homo}
\min_{\sigma_{\text{K}} \geq 0,\sigma_{\text{U}}>0 } 
\mathbb{E}_{S \sim g(s)}
\left[ \frac{S \sigma_{\text{K}}^2 +\sigma_{\text{U}}^2}{N-S} \right] 
 \\
\text{s.t. }
\frac{(N-\bar{C}-1)\sigma_{\text{K}}^2+\sigma_{\text{U}}^2}{((N-\bar{C})\sigma_{\text{K}}^2+\sigma_{\text{U}}^2)^2
	\left( \sigma_{\text{U}}^2\right)
	^2 
} 
\\
\left( 
\left( N-\bar{C}-1\right) \sigma_{\text{K}}^4
+ \left( \sigma_{\text{K}}^2+\sigma_{\text{U}}^2\right)
^2 
\right)
 \leq 
\frac{\varepsilon^2 } {{2\log(2/\delta)} \Delta^2}.
\end{multline}
\end{prop}
\begin{IEEEproof}(Sketch)
Following the i.i.d. distribution of the number of stragglers $S$ at each round, we take (\ref{Equ:n_D}) into $\mathbb{E}_{S \sim g(s)} \mathbb{E}\|\mathbf{n}_D(S) \|$ to rewrite the objective function. 
We let $A=\left|\mathcal{N}_{\text{S}} \cap \mathcal{N}_{\text{C}}  \right| \in [0,\min(C,S)]$ denotes the number of clients from both the colluder set and straggler set.
Based on the DP privacy constraint from Proposition \ref{Prop:general DP}, we consider the worst case, by examining all $C \in [0,\bar{C}]$, $S \in [0,\bar{S}]$ $A \in [0,\min(C,S)]$. 
For example, when $S=0$, all pairwise noise terms are canceled in the aggregated disturbance term $\bm{m}_D$, but the constraint in problem (\ref{Opt:homo}) can still guarantee $(\varepsilon,\delta)$-DP for all clients in set $\mathcal{I}_1$. 
\end{IEEEproof}

Next, we assume $S$ is uniformly distributed in the range $[0,\bar{S}]$, then $\mathbb{E}_{S \sim g(s)}\left[ \frac{S \sigma_{\text{K}}^2 +\sigma_{\text{U}}^2}{N-S}\right] $ takes the form of $\frac{1}{\bar{S}+1} \left(\sum_{s=0}^{\bar{S}} \frac{s \sigma_{\text{K}}^2}{N-s}+ \sum_{s=0}^{\bar{S}} \frac{ \sigma_{\text{U}}^2}{N-s}\right) $.
We then obtain the optimal pair of noise levels $(\sigma_{\text{K}},\sigma_{\text{U}})$ in Example \ref{ex:variance sol-homo}. This could be generalized to any other distributions.
\begin{example}
\label{ex:variance sol-homo}
Consider $S$ is uniformly distributed in the discrete range $[0,\bar{S}]$. 
Let 
\begin{equation*}
\begin{array}{l}
\mu=\frac{\sum_{s=0}^{\bar{S}} \frac{s}{N-s}}{\sum_{s=0}^{\bar{S}} \frac{ 1}{N-s}}, \\
\kappa_4=2 \mu (N-\bar{C})^3-2 \mu (N-\bar{C})^2, \\
\kappa_3=(N-\bar{C})^3-(N-\bar{C})^2+7 \mu (N-\bar{C})^2-6 \mu (N-\bar{C}),\\
\kappa_2=3 (N-\bar{C})^2-3 (N-\bar{C})+9 \mu (N-\bar{C})-6 \mu, \\
\kappa_1=-(N-\bar{C})^2+5 (N-\bar{C})-4+\mu (N-\bar{C})+2 \mu, \\
\kappa_0=-(N-\bar{C})+1+\mu,
\end{array}
\end{equation*}	
then there exists $\gamma_0 \in (0,1)$ to be the minimum positive solution of $G(\gamma)=\kappa_4 \gamma^4+\kappa_3 \gamma^3+\kappa_2 \gamma^2+\kappa_1 \gamma+\kappa_0=0$.
The pair of noise levels
	\label{Equ:sigma_U}
$\sigma_{\text{U}}^*=\frac{\sqrt{
		2\log(\frac{2}{\delta})
		((N-\bar{C}-1) \gamma_0+1) \left((N-\bar{C}-1)\gamma_0^2 +(\gamma_0+1)^2\right)}\Delta}{\varepsilon ((N-\bar{C})\gamma_0+1)} 
$
and
$\sigma_{\text{K}}^*=\sqrt{\gamma_0} \sigma_{\text{U}}^*$
optimally solves problem (\ref{Opt:homo}), i.e., minimize the expected convergence bound from Proposition \ref{Prop:convergence bound} under DP constraints specified in Proposition \ref{Prop:general DP}.
\end{example}

The results of Example \ref{ex:variance sol-homo} are intuitive to understand. With a minimum value of $\sigma_{\text{U}}^2$ for the individual noise $\bm{n}_i$'s, $\sigma_{\text{K}}^2$ can be derived accordingly to add sufficient randomness to local updates.
Such choices work for flexible values of $C \in [0,\bar{C}]$, $S^{(t)} \in [0,\bar{S}]$ and any intersection cases for subsets $\mathcal{N}_{\text{C}}$ and $\mathcal{N}_{\text{S}}^{(t)}$. 
Fig.\ref{fig:theorem results} further illustrates the results in Example \ref{ex:variance sol-homo}.

\begin{itemize}
\item 
\textbf{Impact of colluder set}:
As the maximum number $\bar{C}$ of colluders increases, both $\sigma_{\text{U}}^*$ and $\sigma_{\text{K}}^*$ increases.
$\sigma_\text{U}^*$ is more sensitive to the changes of $\bar{C}$, because the individual noise terms $\bm{n}_i$'s may defend alone against increasing colluders in the accumulated disturbance term $\bm{m}_D$ when $S=0$ (no stragglers). 
In contrast, $\sigma_\text{K}^*$ slightly increases with $\bar{C}$, as $N-\bar{C}-1$ pairwise noise terms $\bm{r}_{ij}$'s remain together in the local disturbance term $\bm{m}_i$ for client $i$.

\item 
\textbf{Impact of straggler set}:
As the maximum number $\bar{S}$ of stragglers increases, $\sigma_{\text{U}}^*$ increases while $\sigma_{\text{K}}^*$ decreases. The reason behind is that it is possible that there are more stragglers in the system with a greater $\bar{S}$, leaving more pairwise terms $\bm{r}_{ij}$'s in the global noise term $\bm{n}_D$. Hence, $\sigma_{\text{K}}^*$ decreases for better accuracy performance, while $\sigma_{\text{U}}^*$ has to increase instead to bound the whole privacy loss.

\item 
\textbf{Impact of privacy budget}:
As the privacy budget $\varepsilon$ increases, both $\sigma_{\text{U}}^*$ and $\sigma_{\text{K}}^*$ decrease under a looser privacy requirement.
\end{itemize}

\section{Experimental Validation}
\label{Sec:experiment}


In this section, we conduct the experimental study  to evaluate the performance of the proposed LightDP-FL scheme and compare with existing schemes under the same DP level. 

\subsection{Experiment Setup }   
We evaluate the performance of the proposed LightDP-FL scheme described in Section \ref{Sec:our scheme} on the CIFAR-10 dataset \cite{krizhevskyCIFAR10} using the ResNet-18 model \cite{he2016deep}. The CIFAR-10 dataset consists of colored photographs categorized into 10 different object classes. We consider a non-i.i.d. data distribution and use FedAvg \cite{mcmahan2017communication} for model aggregation. The experiments are conducted over $T=150$ communication rounds with $N=50$ clients.
We set the batch size and learning rate as 32 and 0.05, respectively.
All results are averaged over 5 trials.
We use test accuracy as our evaluation metric, defined as the ratio of correctly classified images to the total number of images in the test set. 

To demonstrate the efficacy of LightDP-FL, we compare our results with two general baselines:
1) vanilla noise adding with local differential privacy (LDP) algorithm \cite{abadi2016deep,wei_federated_2020}, which requires each client to add local noise and make his local parameter differentially-private before sending it to an untrusted central server; and
2) combination of SMPC with DP scheme, which utilizes SMPC schemes such as homomorphic encryption (HE) \cite{truex2019hybrid}, pairwise encryption (SecAgg) \cite{bonawitz2017practical}, and coded computation \cite{jahani2023swiftagg+}, to ensure the secure transmission of updates from clients to the server. Additionally, a small amount of random noise is added locally to ensure that the aggregated parameters at the server side are DP.

\begin{figure}[!t]
    \centering
    \begin{subfigure}[t]{0.22\textwidth}
        \centering
        \includegraphics[width=\textwidth]{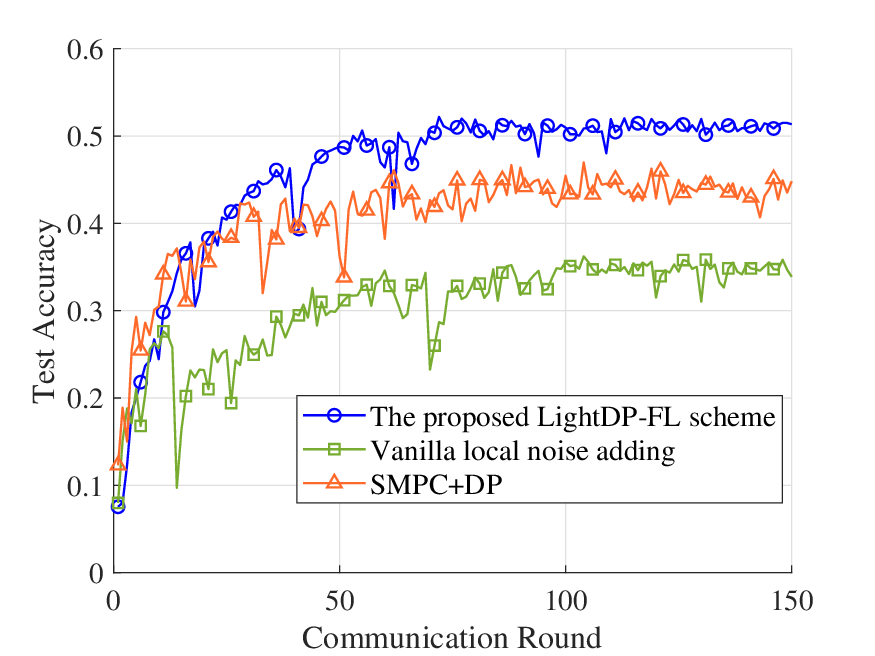}
        \caption{$\epsilon=3$.}
        \label{fig:epsilon3}
    \end{subfigure}
    \hfill
    \begin{subfigure}[t]{0.22\textwidth}
        \centering
        \includegraphics[width=\textwidth]{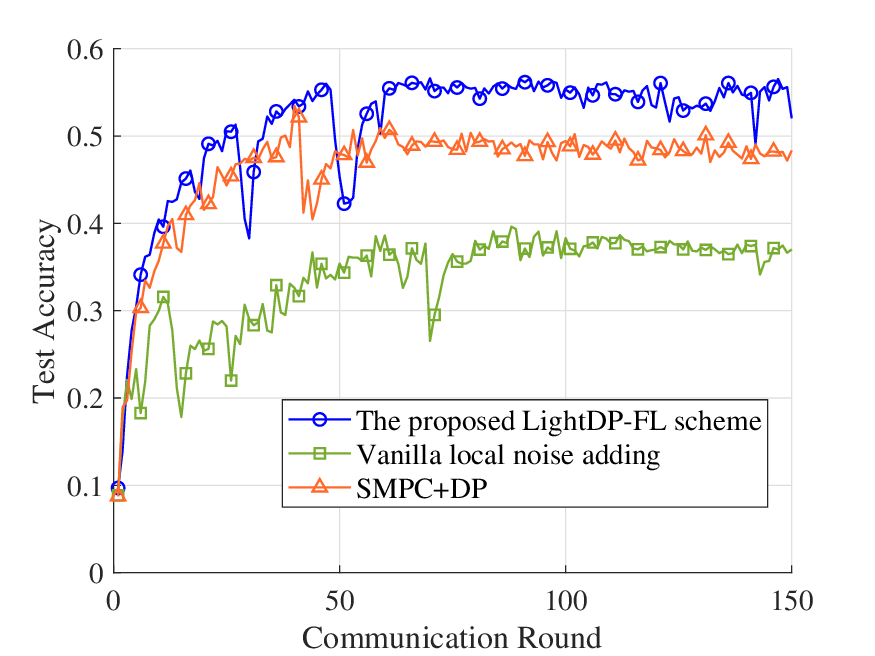}
        \caption{$\epsilon=6$.}
        \label{fig:epsilon6}
    \end{subfigure}
    \hfill
    \begin{subfigure}[t]{0.22\textwidth}
        \centering
        \includegraphics[width=\textwidth]{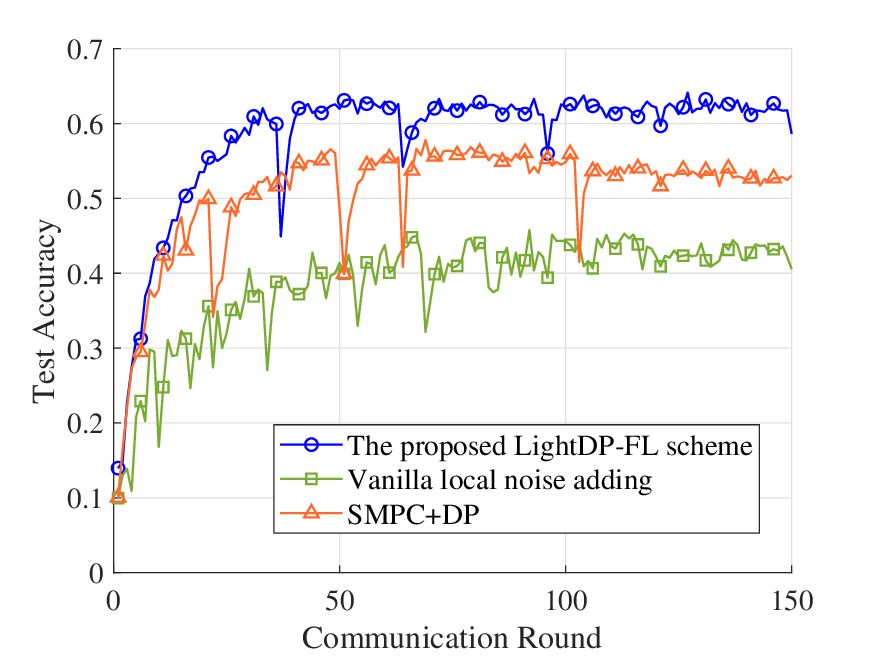}
        \caption{$\epsilon=9$.}
        \label{fig:epsilon9}
    \end{subfigure}
    \caption{Test accuracy with the CIFAR-10 dataset using a ResNet-18 model under LightDP-FL, the SMPC+DP scheme, and the vanilla local noise-adding scheme for different privacy budgets $\varepsilon=3, 6,$ and $9$. We set $N=50$, $\bar{C}=10$, $\bar{S}=10$.}
    \label{fig:conv-epsilon}
\end{figure}

\begin{figure}[!t]
	\centering
	\includegraphics[width=2 in]{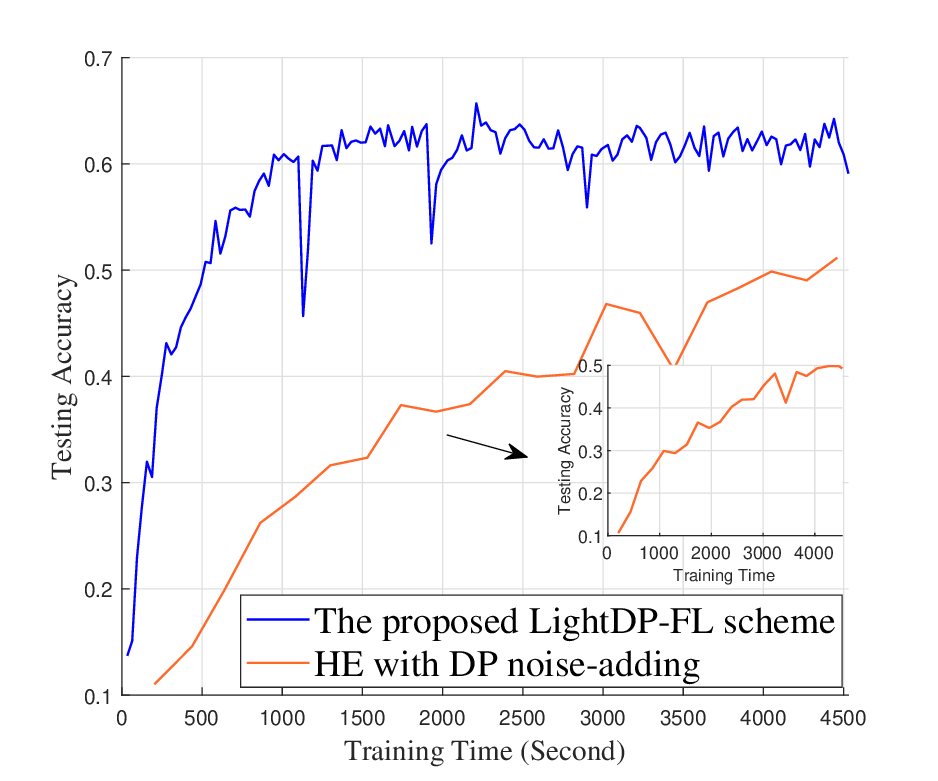}
	        \caption{Performance comparison of LightDP-FL with the combination of HE and DP-noise adding for test accuracy of training a ResNet-18 model on the CIFAR-10 dataset.}
	        \label{Fig:compare_on_CIFAR-10}
\end{figure}

\subsection{Performance Evaluation }

In Fig.~\ref{fig:conv-epsilon}, we compare the test accuracy of the proposed LightDP-FL, with two baselines.
Compared to the vanilla DP noise-adding baseline, which adds much local noise,  the proposed scheme speeds up the training convergence and increases the test accuracy by more than $15\%$.
Compared to the combination of SMPC (e.g., HE \cite{truex2019hybrid} and SecAgg \cite{bonawitz2017practical}) with DP noise-adding, Fig.~\ref{fig:conv-epsilon} shows our scheme still achieves better accuracy performance.
The primary reason for this improvement is that, even with SMPC, the random noise relies on the maximum number of colluders ($\bar{C}$) and stragglers ($\bar{S}$), and must be added in the worst-case scenario, i.e., by a factor of $\frac{1}{N-\bar{C}-\bar{S}}$ of the vanilla noise-adding without SMPC. Otherwise, in the case of fewer numbers of stragglers $S$ than $\bar{S}$, the aggregation at the server could potentially break DP without sufficient noise. 
For example, imagine a scenario with no stragglers, all $N$ clients would add random noise with variance $\frac{\sqrt{1.25 \log(2/\delta)} \Delta}{\varepsilon (N-\bar{C}-\bar{S})}$, where $\frac{\sqrt{1.25 \log(2/\delta)} \Delta}{\varepsilon }$ is the noise variance for the vanilla local DP-noise adding.
This may lead to excessive accumulated noise during the global aggregation. In contrast, our scheme considers the distribution $g(S)$ of $S$ in problem (\ref{Opt:homo}) to more flexibly improve convergence accuracy, as detailed in Section \ref{Subsec:noise opt}. This scheme allows for better performance by adjusting the noise variance based on the actual distribution of the number of stragglers, rather than relying solely on worst-case assumptions.

Further, Fig.~\ref{Fig:compare_on_CIFAR-10} compare these schemes with the overhead and straggler issue more fairly on the training time instead of the training round. We assume a small probability 0.05 of failed transmission for each link in the schemes. SecAgg \cite{bonawitz2017practical} and SwiftAgg+ \cite{jahani2023swiftagg+}, which rely heavily on a large amount of client-to-client transmissions, suffer from the fragile link failure and thus fail to train the model. Thus we cannot show their training performance in Fig.~\ref{Fig:compare_on_CIFAR-10}. 
Specifically, given a total of $\frac{N(N-1)}{2}$ client-to-client links and $N$ client-to-server links, the failure probability of 0.05 results in a much larger number of stragglers compared to the bound $\bar{S}$. In this case, the server actually cannot recover the ground-truth aggregation and fails training.
The HE scheme, on the other hand, suffers from significantly longer computation time per training round, resulting in slower convergence, by at least 10 times compared to LightDP-FL.

\subsection{Trade-off between Privacy and Accuracy}
In Tables~2 and 3, we show the testing accuracy of LightDP-FL with respect to the privacy budget ($\varepsilon$), under different maximum numbers of colluders ($\bar{C}$) and stragglers ($\bar{S}$).  
Note that here we optimize the noise variances according to Example \ref{ex:variance sol-homo}. 

As the privacy budget $\varepsilon$ decreases, the noise variances $\sigma_{\text{K}}^2$ and $\sigma_{\text{U}}^2$ designed increase according to Example \ref{ex:variance sol-homo}. Hence, the convergence error increases according to Proposition \ref{Prop:convergence bound}.
Furthermore, we also observe that under the same privacy budget, the testing accuracy decreases as $\bar{C}$ increases, as greater noise variances (both individually and pairwisely) are needed to provide enough DP guarantee.
The testing accuracy decreases as $\bar{S}$ increases. The reason is that when it is more likely to have a greater number of stragglers, we prefer to lower the variances of pairwise noise terms to keep straggler-resilient and avoid possible accuracy loss.

\begin{table}[]
	\centering
	\caption{Privacy-accuracy tradeoff of the proposed LightDP-FL scheme with different upper bounds of colluders $\bar{C}$ on CIFAR10 dataset, when fixing $\bar{S}=10$.}
	\begin{tabular}{c|cccc}
		\hline
		\multirow{2}{*}{Privacy budget} &
		\multicolumn{4}{l}{Testing accuracy ($\%$)}                                   \\ \cline{2-5} 
		& \multicolumn{1}{l|}{$\bar{C}=10$} & \multicolumn{1}{l|}{$\bar{C}=20$} & \multicolumn{1}{l|}{$\bar{C}=30$} &
		$\bar{C}=40$  \\ \hline
		$\varepsilon=3$
		& \multicolumn{1}{l|}{50.8} & \multicolumn{1}{l|}{49.6} & \multicolumn{1}{l|}{47.5} &
		45.8  \\ \hline
		$\varepsilon=6$
		& \multicolumn{1}{l|}{54.5} & \multicolumn{1}{l|}{52.1} & \multicolumn{1}{l|}{50.6}
		& 49.7 \\ \hline
		$\varepsilon=9$
		& \multicolumn{1}{l|}{61.9} & \multicolumn{1}{l|}{60.6} & \multicolumn{1}{l|}{57.9}
		& 55.3 \\ \hline
	\end{tabular}
	\label{Table:trade-off for C}
\end{table}

\begin{table}[]
	\centering
	\caption{Privacy-accuracy tradeoff of the proposed LightDP-FL with different upper bounds of stragglers $\bar{S}$ on CIFAR10 dataset, when fixing $\bar{C}=15$.}
	\begin{tabular}{c|cccc}
		\hline
		\multirow{2}{*}{Privacy budget} &
		\multicolumn{4}{l}{Testing accuracy ($\%$)}                                   \\ \cline{2-5} 
		& \multicolumn{1}{l|}{$\bar{S}=0$} & \multicolumn{1}{l|}{$\bar{S}=10$} & \multicolumn{1}{l|}{$\bar{S}=20$} & $\bar{S}=30$  \\ \hline
$\varepsilon=3$
& \multicolumn{1}{l|}{51.7} & \multicolumn{1}{l|}{50.2} & \multicolumn{1}{l|}{48.5} & 47.2  \\ \hline
$\varepsilon=6$ & \multicolumn{1}{l|}{53.2} & \multicolumn{1}{l|}{51.7} & \multicolumn{1}{l|}{50.4} & 48.8 \\ \hline
$\varepsilon=9$ & \multicolumn{1}{l|}{55.1} & \multicolumn{1}{l|}{53.4} & \multicolumn{1}{l|}{51.9} & 50.9 \\ \hline
	\end{tabular}
	\label{Table:trade-off for S}
\end{table}

%
%

%
%

\section{Conclusion}
\label{Sec:Conclusion}
In this paper, we presented a lightweight scheme for FL that properly adds both pairwise and individual Gaussian noise to mask local parameters. The proposed LightDP-FL scheme is easily carried out with low communication and computation loads and ensures provable DP. Benefiting from the properly designed small noise terms and only one round-trip transmission per round, our scheme offers robustness to straggling clients. Experiment results show our high accuracy performance, stronger straggler resilience and faster convergence under the same DP level compared to existing baselines.

%


\appendices

\section{Proof of Lemma \ref{Lemma:covariance matrix}}
\label{Proof:Lemma:covariance matrix}
\begin{IEEEproof}
To find the covariance between any two disturbance terms $\bm{m}_i$ and $\bm{m}_j$ ($i,j \in \mathcal{I}_1$ and $i \neq j$), we have
\begin{align*}
Cov(\bm{m}_i,\bm{m}_j) 
= E[(\bm{m}_i- E[\bm{m}_i])(\bm{m}_j - E[\bm{m}_j])]\\
=E[(\bm{m}_i)(\bm{m}_j)].
\end{align*}
Let $\bm{m}_i=\mathbf{A}+\bm{r}_{ij}$ and $\bm{m}_j=\mathbf{B}-\bm{r}_{ij}$, where $\mathbf{A}$ and $\mathbf{B}$ denote the other terms in (\ref{Equ:local disturbance term m i}) except for the dependent $\bm{r}_{ij}$. Notice that $\mathbf{A}$ and $\mathbf{B}$ are independent with each other and with $\bm{r}_{ij}$.
Given that $E[(\bm{m}_i)(\bm{m}_j)]=E[(\mathbf{A}+\bm{r}_{ij})(\mathbf{B}-\bm{r}_{ij})]=E[\mathbf{A} \mathbf{B}]
	-E[\mathbf{A} \bm{r}_{ij}]
	+E[\mathbf{B} \bm{r}_{ij}]
	-E[( \bm{r}_{ij})^2]
	=-E[\bm{r}_{ij}^2]
	$, we have
\begin{multline*}
Cov(\bm{m}_i,\bm{m}_j) 
=C_{ij}=C_{ji}
=-E[\bm{r}_{ij}^2] \\
=-Var(\bm{r}_{ij})-(E[\bm{r}_{ij}])^2
=-\sigma_{ij}^2
\end{multline*}
and
\begin{equation*}
Cov(\bm{m}_i,\bm{m}_i) =C_{ii}
=Var(\bm{m}_i)
=\sum_{j \neq i, j\in \mathcal{N}\setminus\mathcal{N}_{\text{C}}} \sigma_{ij}^2+\sigma_i^2.
\end{equation*}

We then have the symmetric covariance matrix for all disturbance terms $\bm{m}$ in (\ref{Matrix:C_m}) and the multivariate Gaussian distribution with the zero mean of $\bm{m}$.	
\end{IEEEproof}

\section{Proof of Lemma \ref{Lemma:write privacy loss as Gaussian}}
\label{Proof:Lemma:write privacy loss as Gaussian}
\begin{IEEEproof}
Based on two adjacent datasets $\mathcal{D}_i$ and $\mathcal{D}_i'$ for a single client $i \in \mathcal{I}_1$, we consider two adjacent matrices
$\mathbf{X}^t=\{\bm{x}_1,\cdots,\bm{x}_i,\cdots,\bm{x}_{|\mathcal{I}_1|} \}$ and $\hat{\mathbf{X}}^t=\{\bm{x}_1,\cdots,\bm{x}_i+\bm{v},\cdots,\bm{x}_{|\mathcal{I}_1|}\}$ to denote the resulting adjacent observations.
Then we rewrite the privacy loss in (\ref{Equ: privacy loss-initial ratio form}) as
\begin{equation*}
\mathcal{L}_i
=
\ln
\frac{
	\exp \left\{-\frac{\mathbf{X}^t \mathbf{C}_{\bm{m}}^{-1}\mathbf{X}}{2}\right\}
}
{
	\exp \left\{-\frac{\hat{\mathbf{X}}^t \mathbf{C}_{\bm{m}}^{-1}\hat{\mathbf{X}}}{2}\right\}
} 
=
-\frac{1}{2} (  
\mathbf{X}^t \mathbf{C}_{\bm{m}}^{-1}\mathbf{X}
-
\hat{\mathbf{X}}^t \mathbf{C}_{\bm{m}}^{-1}\hat{\mathbf{X}}
).
\end{equation*}
Accounting for the changes in $\bm{x}_i$ by $\bm{v}$, we have
\begin{equation*}
\begin{aligned}
\mathcal{L}_i&=
\bm{v}^t \sum_{j \in \mathcal{I}_1:j \neq i} C_{i j}^{-1} \bm{x}_j 
+\frac{1}{2} C_{i i}^{-1}\left(2 \bm{v}^t \bm{x}_i+\|\bm{v}\|_2^2\right) \\
&=
\bm{v}^t
\sum_{j \in \mathcal{I}_1} C_{i j}^{-1} \bm{x}_j 
+\frac{C_{i i}^{-1}}{2} \|\bm{v}\|_2^2,
\end{aligned}
\end{equation*}
where $C_{i j}^{-1}$ is the $(i,j)$ element of the inverse of the covariance matrix, $\mathbf{C}_{\bm{m}}^{-1}$, $\|\bm{v}\|_2^2$ is the squared $L2$ norm of vector $\bm{v}$.


Since $\bm{x}_i$ follows a Gaussian distribution, the privacy loss random variable $\mathcal{L}_i$ is also Gaussian and distributed with mean $\frac{C_{i i}^{-1}}{2} \|\bm{v}\|_2^2$ and variance
\begin{equation*}
\begin{aligned}
&\left( 
\sum_{j \in \mathcal{I}_1} (C_{i j}^{-1})^2 Var(\bm{x}_j)
\right) 
\|\bm{v}\|_2^2 \\
=&
\left[
\sum_{j \in \mathcal{I}_1} (C_{i j}^{-1})^2
\left( 
\sum_{k \neq j, k \in \mathcal{N} \setminus \mathcal{N}_{\text{C}}} \sigma_{j k}^2 +\sigma_j^2
\right)   
\right] 
\|\bm{v}\|_2^2 \\
=&
\left[
\sum_{j \in \mathcal{I}_1} (C_{i j}^{-1})^2
C_{j j}   
\right] 
\|\bm{v}\|_2^2.
\end{aligned}
\end{equation*}
\end{IEEEproof}

\section{Proof of Proposition \ref{Prop:general DP}}
\label{Proof:Prop:general DP}
\begin{IEEEproof}
Let $Z \sim \mathcal{N}(0,1)$, then we rewrite the privacy loss random variable according to Lemma \ref{Lemma:write privacy loss as Gaussian} as Gaussian:
\begin{align*}
	\sqrt{\sum_{j \in \mathcal{I}_1} (C_{i j}^{-1})^2
		C_{j j}  } \|\bm{\omega}_i(\mathcal{D}_i)-\bm{\omega}_i(\mathcal{D}_i')\|_2 Z
		\\ 
		+
		\frac{C_{ii}^{-1}}{2}
		\|\bm{\omega}_i(\mathcal{D}_i)-\bm{\omega}_i(\mathcal{D}_i')\|_2^2.
	\end{align*}
The probability that it exceeds $\varepsilon$ (i.e., $\operatorname{Pr}\left( 
\left|
\sqrt{\sum_{j \in \mathcal{I}_1} (C_{i j}^{-1})^2
	C_{j j} } \|\bm{\omega}_i(\mathcal{D}_i)-\bm{\omega}_i(\mathcal{D}_i')\|_2 Z 
\right.\right. \\
\left.\left.
+
\frac{1}{2}C_{ii}^{-1}
\|\bm{\omega}_i(\mathcal{D}_i)-\bm{\omega}_i(\mathcal{D}_i')\|_2^2
\right| \geq \varepsilon  \right)$) can be rewritten as
$\operatorname{Pr}
		\left( 
		|Z| \geq
\frac{\varepsilon }{
\sqrt{\sum_{j \in \mathcal{I}_1} (C_{i j}^{-1})^2
		C_{j j}  } \|\bm{\omega}_i(\mathcal{D}_i)-\bm{\omega}_i(\mathcal{D}_i')\|_2
}
\right. \\
\left.
		-
		\frac{C_{ii}^{-1} \|\bm{\omega}_i(\mathcal{D}_i)-\bm{\omega}_i(\mathcal{D}_i')\|_2}{
			2 \sqrt{\sum_{j \in \mathcal{I}_1} (C_{i j}^{-1})^2
					C_{j j}   }
		}
		\right)$
At this point, we will be a bit informal and drop the latter term for the sake of presentation, which is common in DP proofs. 
%
Based on the standard Gaussian tail bounds, e.g., $\operatorname{Pr}
\left( 	|Z| \geq v	\right) \leq 2 \exp(-\frac{v^2}{2})$, we have $\operatorname{Pr}\left(|Z| \geq \sqrt{2\log(2/\delta)} \right) \leq \delta $. Therefore, the sufficient condition of all clients' $(\varepsilon,\delta)$-DP privacy is
\begin{multline*}
\operatorname{Pr}
		\left( 
		|Z| \geq
\frac{\varepsilon }{
\sqrt{\sum_{j \in \mathcal{I}_1} (C_{i j}^{-1})^2
		C_{j j}  } \|\bm{\omega}_i(\mathcal{D}_i)-\bm{\omega}_i(\mathcal{D}_i')\|_2
}
		\right)  \\
\leq
\operatorname{Pr}\left(|Z| \geq \sqrt{2\log(2/\delta)} \right)		
		\leq \delta.
\end{multline*}
Equivalently, we have
$\frac{\varepsilon }{
\sqrt{\sum_{j \in \mathcal{I}_1} (C_{i j}^{-1})^2
		C_{j j}   } \|\bm{\omega}_i(\mathcal{D}_i)-\bm{\omega}_i(\mathcal{D}_i')\|_2
}
\geq 
\sqrt{2\log(2/\delta)}$.
%
\end{IEEEproof}

\bibliographystyle{ieeetr}
\bibliography{Infocom25Ref}

\end{document}